\documentclass[showpacs,prb,final,superscriptaddress,twocolumn]{revtex4-1}
\usepackage{graphicx}
\usepackage{amssymb}
\usepackage{marvosym}
\usepackage{latexsym}
\usepackage[usenames]{color}
\usepackage{float}
\usepackage{array}
\usepackage{verbatim} 
\usepackage{natbib} 
\usepackage{hyperref}

\begin{document}

\title{The Effect of Nickel Substitution on Magnetism in the Layered van der Waals Ferromagnet Fe$_3$GeTe$_2$}

\author{Gil Drachuck}
\affiliation{Department of Physics and Astronomy, Iowa State University, Ames, Iowa 50011, USA}
\affiliation{Ames Laboratory, U. S. DOE, Iowa State University, Ames, Iowa 50011, USA}
\author{Zaher Salman}
\affiliation{Laboratory for Muon Spin Spectroscopy, Paul Scherrer Institut, CH-5232 Villigen PSI, Switzerland}
\author{Morgan W. Masters}
\affiliation{Department of Physics and Astronomy, Iowa State University, Ames, Iowa 50011, USA}
\author{Valentin Taufour}
\affiliation{Department of Physics and Astronomy, Iowa State University, Ames, Iowa 50011, USA}
\affiliation{Department of Physics, University of California Davis, Davis, CA 95616, USA}
\author{Tej N. Lamichhane}
\affiliation{Department of Physics and Astronomy, Iowa State University, Ames, Iowa 50011, USA}
\affiliation{Ames Laboratory, U. S. DOE, Iowa State University, Ames, Iowa 50011, USA}
\author{Qisheng Lin}
\affiliation{Ames Laboratory, U. S. DOE, Iowa State University, Ames, Iowa 50011, USA}
\affiliation{Department of Chemistry, Iowa State University, Ames, Iowa 50011, USA}
\author{Warren E. Straszheim}
\affiliation{Ames Laboratory, U. S. DOE, Iowa State University, Ames, Iowa 50011, USA}
\author{Sergey L. Bud'ko}
\affiliation{Department of Physics and Astronomy, Iowa State University, Ames, Iowa 50011, USA}
\affiliation{Ames Laboratory, U. S. DOE, Iowa State University, Ames, Iowa 50011, USA}
\author{Paul C. Canfield}
\affiliation{Department of Physics and Astronomy, Iowa State University, Ames, Iowa 50011, USA}
\affiliation{Ames Laboratory, U. S. DOE, Iowa State University, Ames, Iowa 50011, USA}
\date{\today }

\begin{abstract}
We have grown a series of nickel substituted single crystals of the layered ferromagnet (FM) Fe$_3$GeTe$_2$. The large single crystalline samples of (Fe$_{1-x}$Ni$_x$)$_3$GeTe$_2$ with $x = 0-0.84$ were characterized with single crystal X-ray diffraction, magnetic susceptibility, electrical resistance and muon spin spectroscopy. We find Fe can be continuously substituted with Ni with only minor structural variation. In addition, FM order is suppressed from $T_\mathrm{C}=212$~K for $x=0$ down to $T_\mathrm{C}=50$~K for $x=0.3$, which is accompanied with a strong suppression of saturated and effective moment, and Curie-Weiss temperature. Beyond $x=0.3$, the FM order is continuously smeared into a FM cluster glass phase, with a nearly full magnetic volume fraction. We attribute the observed change in the nature of magnetic order to the intrinsically disordered structure of Fe$_3$GeTe$_2$ and subsequent dilution effects from the Ni substitution. 
\end{abstract}

\maketitle

\section{Introduction}

During the last two decades, a large number of ferromagnetic (FM) metals with low Curie temperature (T$_\mathrm{C}$) have been discovered. In these materials, mechanical pressure, magnetic field or chemical substitution can tune the system across a paramagnetic-ferromagnetic (PM-FM) quantum phase transition (QPT) and often reveal peculiar magnetic ground states. Applying pressure in clean systems such as URhGe~\cite{Levy2007}, ZrZn$_2$~\cite{Uhlarz2004},UGe$_2$~\cite{Saxena2000,Taufour2010}, LaCrGe$_3$~\cite{TaufourPRL2016,Kaluarachchi2017} and others, suppresses FM and drives the system towards a quantum phase transition (QPT), where peculiarly the nature of the FM-PM transitions changes before being completely suppressed~\cite{Brando2016}. For example, the FM transition can become first order, develop a spin-density wave order and exhibit tricritical wings in a magnetic field, and in some cases even  develop unconventional superconductivity. In contrast, in FM systems where either intrinsic disorder is present or FM is suppressed with chemical substitution (Sr$_{1-x}$Ca$_x$RuO$_3$~\cite{Demk2012}, UNi$_{1-x}$Co$_x$Si$_2$\cite{PikulPRB2012}, U$_{1-x}$Th$_x$NiSi$_2$~\cite{Pikul2012}), the suppression of FM order often results in a smeared QPT after which the system goes into a short range spin glass freezing. In other disordered systems such as CePd$_{1-x}$Rh$_x$ and Ni$_{1-x}$V$_x$ were shown to exhibit a quantum Griffiths region near the FM QCP.~\cite{Westerkamp2009, Kassis2010} 

Another, recent, itinerant FM system is Fe$_3$GeTe$_2$, with a layered van der Waals structure and a T$_\mathrm{C}$ = 220~K~\cite{Deiseroth2006}. It crystallizes into an hexagonal structure ($P6_3/mmc$, 194) and can be grown in a single crystalline form~\cite{Chen2013}. The structure of Fe$_3$GeTe$_2$ is intrinsically disordered as it prefers to form with the Fe2 crystallographic site partially occupied with an occupancy of 0.85. It has been shown that in Fe$_{3-y}$GeTe$_2$, FM order is rapidly suppressed when synthesized with intentional Fe deficiencies. However, Fe-deficient samples with $y>0.1$ have proven difficult to synthesize~\cite{May2016}. Nevertheless, nearly isostructural non-magnetic analog, Ni$_3$GeTe$_2$ ($P6_3/mmc$, 194), which differs from Fe$_3$GeTe$_2$ by an interstitial, partially occupied Ni3 site~\cite{Deiseroth2006,Stahl2016}. This structural similarity allows for a continuous substitution between Fe and Ni, without significantly changing the structural properties. Figure~\ref{Figure1Stucture}(b) shows the structure of both compounds. Since Ni is non-magnetic in this structure, substitution of Fe provides an excellent opportunity to study the effect of a dilution on the FM ground state in Fe$_3$GeTe$_2$. 

In this work, large single crystalline samples of (Fe$_{1-x}$Ni$_x$)$_3$GeTe$_2$ with $x = 0-0.84$ have been grown using the high temperature solution growth technique, and their structural and magnetic properties were investigated with bulk and local probe measurements. We find that, with increasing Ni content, long-range FM order is suppressed continuously and is smeared into a spin-glass phase, with a nearly full magnetic volume fraction. We attribute this to (i) growing disorder from alloying of Ni on the Fe1/Fe2 site and the introduction of a third interstitial Ni3 site, and (ii) diluting magnetic Fe with non-magnetic Ni.  


\section{Experimental Methods}
\label{Experimental}

\begin{figure*}[ht]
	\begin{center}
		\includegraphics[width=172mm]{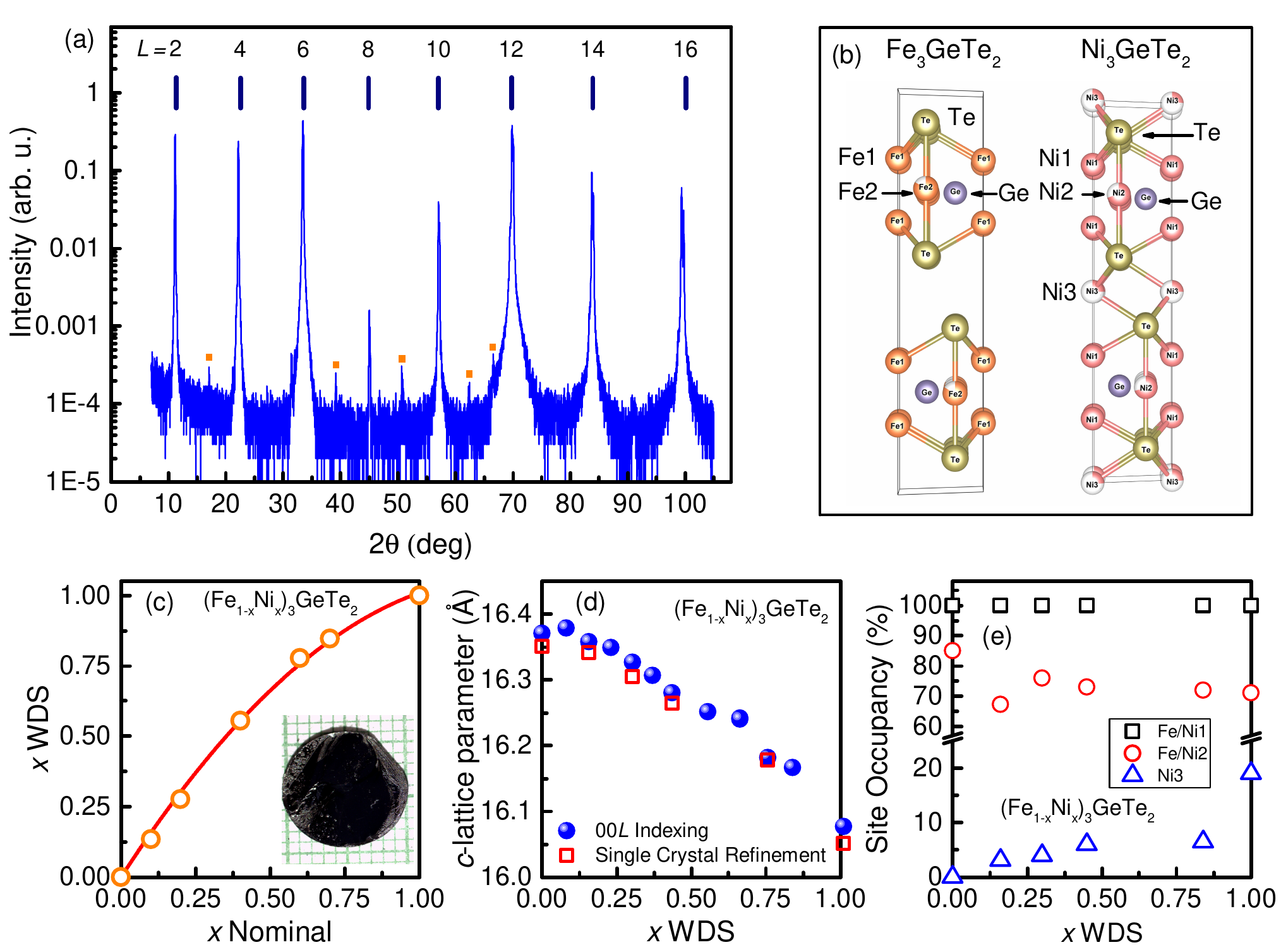}
	\end{center}
	\caption{(a) x-ray diffraction spectrum measured on a single crystal showing (00\textit{L})
		diffraction peaks. Orange squares denote diffraction peaks from secondary phases and/or flux droplets on the surface of the crystal remnant post growth.  (b)  The structure of the end members,  Fe$_3$GeTe$_2$ and Ni$_3$GeTe$_2$. The Shading of the spheres represent the occupancy of each site, for example, the Fe2 site in Fe$_3$GeTe$_2$ has reported 0.85 site occupancy. (c) Ni substitution ($x$ WDS) determined by WDS vs. the nominal Ni composition of the (Fe$_{1-x}$Ni$_x$)$_3$GeTe$_2$ crystals. The red line is a second order polynomial fit to the data ($x_{\mathrm{WDS}} = 1.64(5)x_{\mathrm{nominal}}-0.63(6)x^2_{\mathrm{nominal}}$). Inset: a picture of a crucible limited crystal of a crystal with $x_{\mathrm{WDS}} = 0.16$. (d) \textit{c}-lattice parameter vs. Ni substitution inferred from (00\textit{L}) refinement (full spheres) and from full single crystal refinement (open squares). (e) Refined Fe/Ni site occupancy vs $x_{\mathrm{WDS}}$.  }
	\label{Figure1Stucture}
\end{figure*}

\begin{figure*}[t]
	\begin{center}
		\includegraphics[width=172mm]{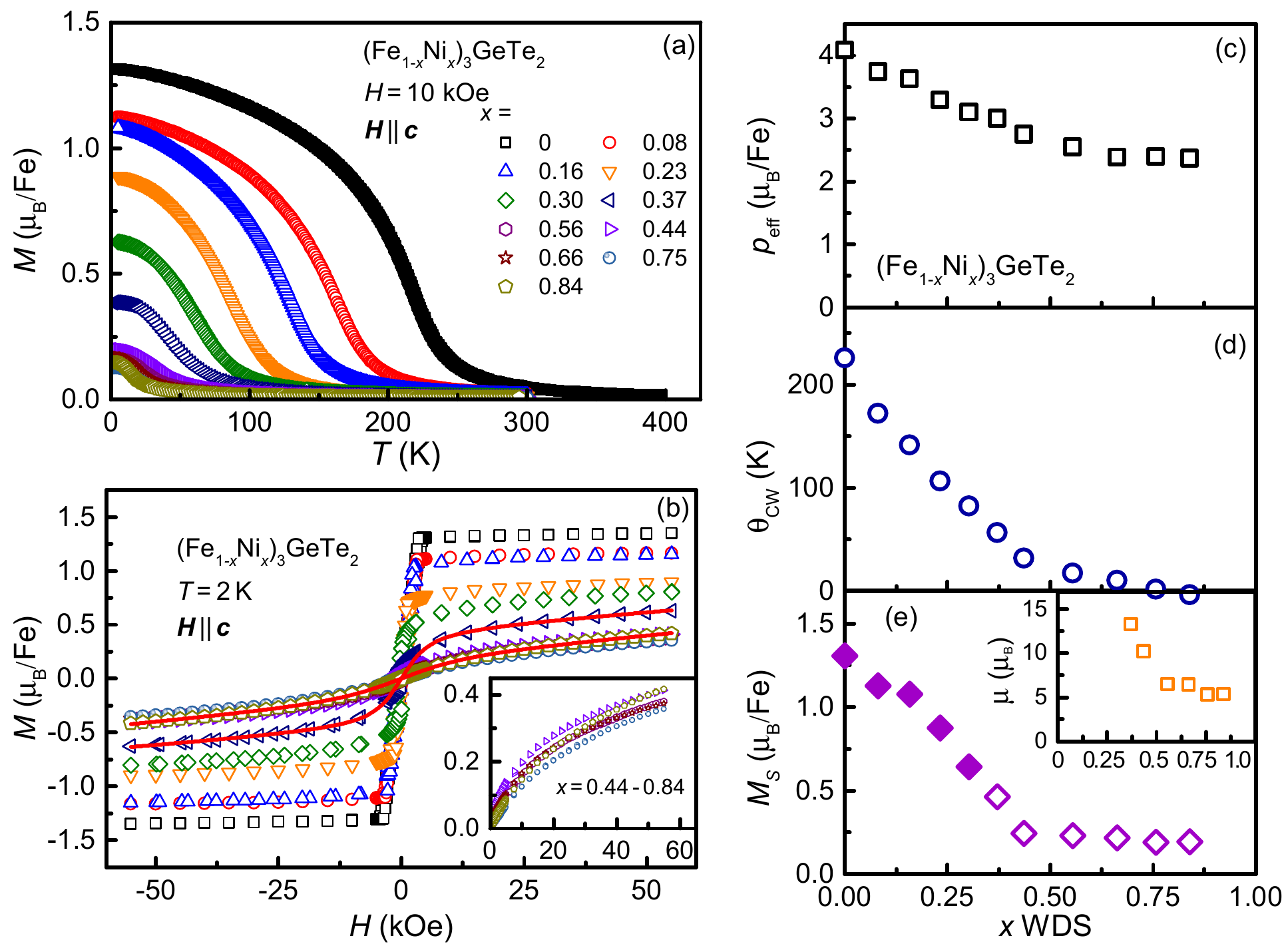}
	\end{center}
	\caption{(a) Magnetic moment per Fe vs temperature measured on (Fe$_{1-x}$Ni$_x$)$_3$GeTe$_2$ along the \textit{c}-axis at $H=$~10~kOe for nickel substitution ranging between $x = 0$ and $x = 0.84$. (b) Magnetization isotherms measured at $T=2$~K along the \textit{c}-axis. The solid red lines are fits to Eq.~\ref{Eq1}. Inset: Blow-up of the magnetization isotherms measured on the non-FM samples ($x=0.44-0.84$).  (c) The inferred effective moment per Fe vs Ni substitution.  (d) Inferred CW temperature vs Ni substitution. (e) Saturated Magnetic moment per Fe vs Ni substitution. Open data points were determined from the fit Langevin equation described in Eq.~\ref{Eq1}. Inset: The average moment per cluster, $\mu$, determined from the fitting the $M(H)$ data to Eq.~\ref{Eq1}.}
	\label{Figure2HighFied}
\end{figure*}

\begin{figure}[t]
	\begin{center}
		\includegraphics[width=86mm]{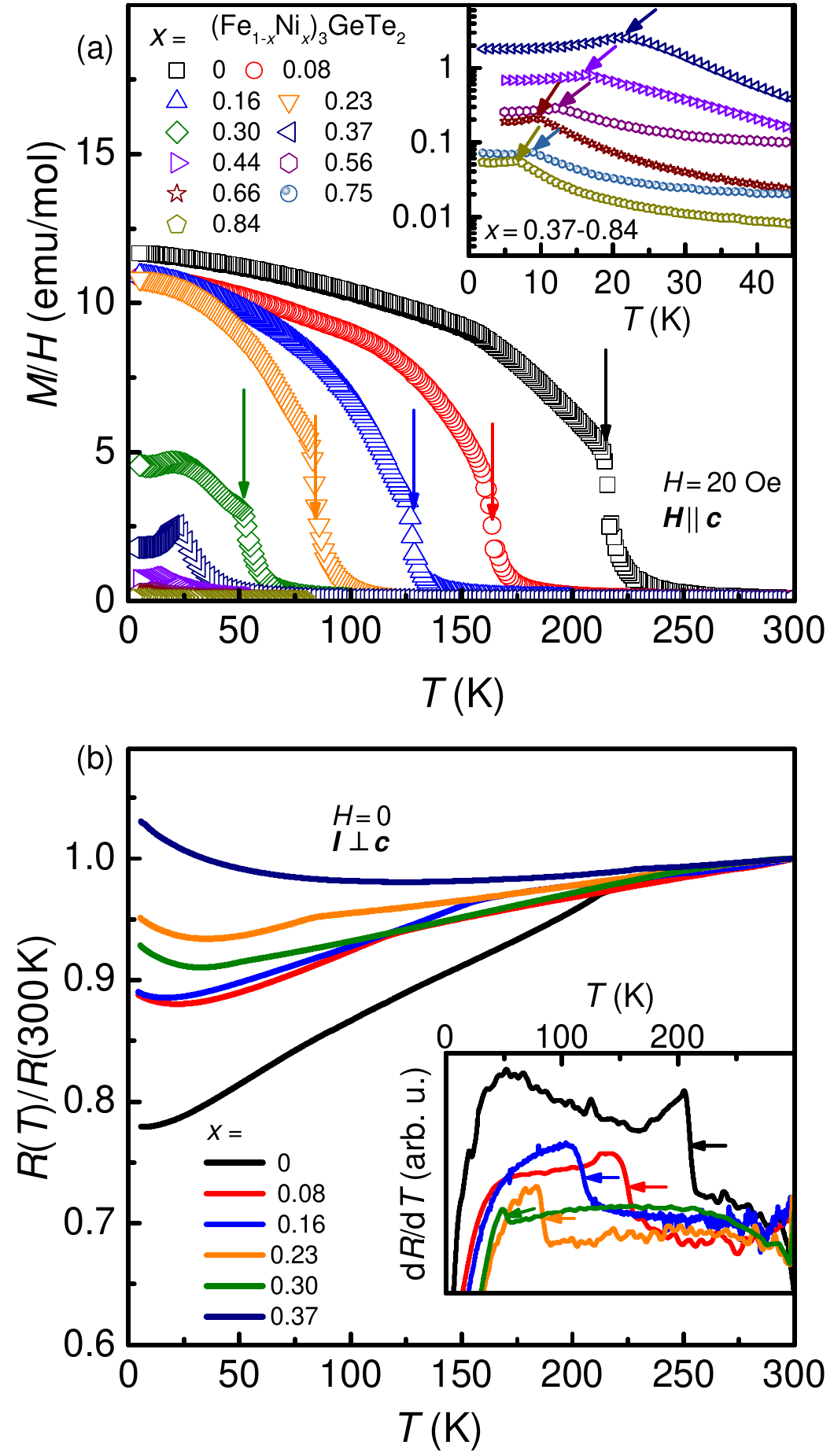}
	\end{center}
	\caption{(a) $M(T)$/$H$ measured at $H=20$~Oe for all compositions ($x=0-0.84$). The arrows correspond to the anomaly observed in the resistance measurement which indicates T$_C$ for the FM samples. Inset: Blow-up of the low temperature  $M(T)$/$H$ curves for samples with $x=0.37-0.84$ on a semi-log plot. (b) Normalized  zero-field in-plane resistance vs temperature for the FM samples of (Fe$_{1-x}$Ni$_x$)$_3$GeTe$_2$ with $x=0-0.3$ and of the non-FM $x=0.37$ sample. Inset: d$R$/d\textit{T} showing a clear anomaly at T$_C$ as marked by the arrows. }
	\label{Figure3LowField}
\end{figure}


Large single crystalline samples of (Fe$_{1-x}$Ni$_x$)$_3$GeTe$_2$ with nickel substitution ranging between $x = 0$ and $x = 0.84$ were grown out of a high-temperature solution rich in Te~\cite{CanfieldFisher2001,CanfieldGrowth}. Powders of Fe and Ni, and pieces of elemental Ge and Te were mixed in molar ratios of (Fe,Ni)$_{0.38}$Te$_{0.56}$Ge$_{0.06}$. The elements were loaded into the bottom 2~ml alumina crucible of a Canfield Crucible Set (CCS)~\cite{CanfieldFrit}, and sealed in amorphous silica ampules under a partial argon atmosphere. The ampules were heated to 460~$^\circ$C in 6 hours and held there for 6 hours, in order to allow the tellurium and iron powder to react, mitigating the risk of rapid ampule disassembly  upon further heating. Subsequently, the ampules were heated over 10 hours to 1000~$^\circ$C and held for 2 additional hours, then heated to 1180~$^\circ$C over 2 hours and held for 3 hours.  The ampules were then slowly cooled, over 60-100 hours to 750~$^\circ$C. At that point, the excess molten Te-rich solution was decanted by a modified centrifuge.~\cite{CanfieldGrowth,CanfieldFrit} In some cases, remnant, trapped flux was found to be enclosed between layers of (Fe$_{1-x}$Ni$_x$)$_3$GeTe$_2$. 

(Fe$_{1-x}$Ni$_x$)$_3$GeTe$_2$ grow as as mirror-like, metallic, micaceous plates with the crystallographic \textit{c}-axis perpendicular to the plate surface  with dimensions ranging from $5\times5\times1$ mm$^3$ up to crucible limited crystals (see inset of Fig~\ref{Figure1Stucture}(c)). They are malleable, and not amenable to grinding for powder x-ray diffraction (XRD) measurements. Instead, XRD from the surface of single crystals were carried out using a Rigaku MiniFlex II powder diffractometer with a Cu K$_{\alpha}$ source and a graphite monochromator in front of the detector~\cite{Jesche2016}. In addition, Single crystal X-ray diffraction intensity data for (Fe$_{1-x}$Ni$_x$)$_3$GeTe$_2$ crystals were collected at room temperature using a Bruker SMART APEX II diffractometer (Mo K$\alpha$ radiation, $\lambda = 0.71073\AA$). Data reduction, integration, unit cell refinements, and absorption corrections were done with the aid of subprograms in APEX2.~\cite{Bruker1,Bruker2} Space group determination, Fourier Synthesis, and full-matrix least-squares refinements on $F^2$ were carried out by in SHELXTL 6.1.~\cite{Bruker3}. The actual composition of the crystals was determined using Wavelength-Dispersive Spectrometry (WDS). 

Temperature and field dependent magnetization measurements were carried out using a Quantum Design Magnetic Property Measurement System (MPMS), superconducting quantum interference device (SQUID) magnetometer (\textit{T} = 1.8 - 300 K, \textit{H}$_{max}$ = 55 kOe). The samples where mounted between two strips of Teflon tape suspended over the edges of two internal straws inserted into an external straw. The magnetization measurements were performed along the magnetic easy axis of Fe$_3$GeTe$_2$ with $H \parallel c$~\cite{Chen2013} .The \textit{c}-axis of the crystals was aligned within 5 degrees of accuracy for the magnetization measurements. Given that the signal of the FM samples was significantly larger than that of the addendum, the magnetization data were not corrected for addendum contribution. 

Electrical resistance was measured using a "Lakeshore Model 370/372" AC resistance bridge in a 4-point probe measurement configuration, in a Janis Research SHI-950T 4 Kelvin Closed Cycle Refrigerator. All resistance measurements were performed with the electrical current $I \perp c$. The uncertainty in determination of the transition temperatures was determined by half width at half maximum in d$M$/d\textit{T} and/or d$R$/d\textit{T}. The error bars due to mass uncertainty and different ranges of CW fit are about 2\% for effective moment and 10\% for paramagnetic Curie-Weiss (CW) temperatures, $\theta_{\mathrm{CW}}$. The uncertainty in the saturated moment value is estimated to be about 2\% as well.

Muon spin relaxation and rotation ($\mu$SR) measurements were performed on the GPS spectrometer at the Paul Scherrer Institute in Switzerland. In our $\mu$SR measurements, the sample was suspended on a Kapton mylar tape in a gas flow cryostat, which allows measurements between 1.6~K and 300~K. The $\mu$SR measurements were performed in the transverse field geometry (TF-$\mu$SR), where an external magnetic field is applied perpendicular to the initial spin polarization.

\section{Results}
\label{Results}

\subsection{Structure and composition}

Figure~\ref{Figure1Stucture} (a) presents x-ray diffraction data of the $\{00\textit{L}\}$ reciprocal planes, collected from a single crystal with $x=0.3$.  The $\{00\textit{L}\}$ reflections with \textit{L} = \textit{even} for all $L\leqslant16$ were identified and indexed. A few non-indexed reflections (marked by orange squares) are apparent in the XRD spectra.  These are a result of a secondary phase inclusions or residual flux from the crystal growth process. Since the secondary reflection are small and incomplete, determination for the secondary phases was infeasible. Compositional analysis was performed on several samples in which the Ni substitution level \textit{x} was determined using WDS. The actual Ni content shown in Fig.~\ref{Figure1Stucture}(c), $x_\mathrm{WDS}$, was found to deviate from the nominal melt composition. However, it follows a quadratic relation, $x_{\mathrm{WDS}} = 1.64(5)x_{\mathrm{nominal}}-0.63(6)x_{\mathrm{nominal}}^2$, which was obtained from the fit to the WDS data (red solid line). This relation was used to determine the Ni content of all samples presented in this work and is simply refereed to as \textit{x}.

The \textit{c}-lattice parameters are presented in Fig.~\ref{Figure1Stucture}(d). They were determined for all Ni compositions from the (00\textit{L}) reflections, according to the procedure described in Ref.~\onlinecite{Jesche2016} (blue spheres). In addition, full single crystal refinement was performed for selected compositions. The refined \textit{c}-lattice parameters is depicted by the red empty squares. Both methods for inferring the \textit{c}-lattice parameters are in good agreement. It is worth noting that although Fe$_3$GeTe$_2$ and Ni$_3$GeTe$_2$ do not share the exact same crystal structure, the \textit{c}-lattice parameter qualitatively follows Vegard's law when the Fe/Ni ratio is varied continuously. 

Figure~\ref{Figure1Stucture}(e) depicts the transition metal (Fe,Ni) site occupancy as a function of Ni substitution refined from single crystal diffraction. The data suggest that Fe/Ni1 site is fully occupied regardless of the Ni content. The Fe/Ni2 site is partly occupied for for all compositions, starting from 85\% for the parent compound, decreasing to 70\% and later stays constant above $x=0.2$. The unique Ni3 site is partly occupied in the Ni$_3$GeTe$_2$ compound at 20\% and is unoccupied in Fe$_3$GeTe$_2$ as previously reported~\cite{Deiseroth2006}. With increasing Ni substitution up to x=0.84, the Ni3 site occupancy monotonically increases and saturates at 6\%. 

\subsection{Bulk Measurements}

In Figure~\ref{Figure2HighFied}(a) the magnetic moment per Fe vs. temperature of the (Fe$_{1-x}$Ni$_x$)$_3$GeTe$_2$ system, measured with a magnetic field $H = 10$~kOe applied along the \textit{c}-axis, is presented for crystals with $x=0-0.84$. The magnetization data show a ferromagnetic transition temperature close to $T=220$~K for the Fe$_3$GeTe$_2$ sample ($x=0$, black squares), consistent with previous reports for both FM ordering temperature and the average size of the ordered moment per Fe~\cite{Chen2013,May2016}. For the Ni substituted samples, both the magnetic transition temperature and the size of the ordered moment per Fe are rapidly suppressed with increasing Ni content up to $x=0.37$. For $x=0.44$ and beyond, the saturated moment per Fe is nearly constant.

Figure~\ref{Figure2HighFied}(b) shows the magnetization isotherms, measured at $T=2$~K, with $H \parallel c$ for crystals with $x=0-0.84$. The data for samples with $x=0-0.3$ is consistent with FM order showing a rapid rise followed by a saturation of the magnetization. For $x\geq0.37$ the rapid increase of the $M(H)$ curves becomes more gradual, as clearly demonstrated in the inset of Fig.~\ref{Figure2HighFied}(b). The shape of $M(H)$ curves  resembles those observed in cluster glasses~\cite{Feng2001,Vasundhara2008} which can be best described by a modified Langevin function represented by
\begin{equation}\label{Eq1}
M\left( H \right) = {M_s}L\left( {{{\mu H} \mathord{\left/
			{\vphantom {{\mu H} {{k_B}T}}} \right.
			\kern-\nulldelimiterspace} {{k_B}T}}} \right) + \chi H
\end{equation}
and was used to fit the $M(H)$ curves for $x\geq0.37$.  Here $\mu$ is the average moment per cluster,  $L\left( x \right) = \coth (x) - {1 \mathord{\left/{\vphantom {1 x}} \right.\kern-\nulldelimiterspace} x}$ is the Langevin function and $M_s$ is the saturation magnetization and $\chi$ is the PM susceptibility. Representative fit curves to the $x=0.37$ and $0.84$ data are shown as solid red lines. Interestingly, the average moment per cluster for the $x = 0.37$ was found to be $13\mu_B$, which monotonically decreases to $5 \mu_B$ for $x = 0.84$ as shown in the inset of Fig.~\ref{Figure2HighFied}(e).

The effective moment per Fe ($p_{\mathrm{eff}}$), CW temperature ($\theta_{\mathrm{CW}}$) were determined from the temperature dependent measurements [Fig.\ref{Figure2HighFied}(a)]. $p_{\mathrm{eff}}$ vs. $x$ is shown in Fig.~\ref{Figure2HighFied}(c) was obtained by fitting the data in Fig.~\ref{Figure2HighFied}(a) to a CW-law above the magnetic transition temperature.  $p_{\mathrm{eff}}$ decreases monotonically with increasing Ni substitution, and exhibits a change in slope around $x=0.44$.  $\theta_{\mathrm{CW}}$ shown in Fig.~\ref{Figure2HighFied}(b) follows a similar trend, decreasing with increasing Ni substitution and showing a change in slope around $x=0.44$. Up to $x=0.37$, $\theta_{\mathrm{CW}}$ is comparable to the FM ordering temperature. Beyond $x=0.44$, it decreases slowly and becomes negative for  $x=0.84$, suggesting a change in the nature of magnetic correlations as a function of Ni substitution.

The value of $M_s$ is shown in Fig.~\ref{Figure2HighFied}(e). For samples with $x=0-0.3$ it determined from the intercept of a linear fit to the for $H>10$~kOe data with the $H=0$ axis in Fig.~\ref{Figure2HighFied}(b) (solid symbols). For sample with $x \geq 0.37$,  $M_s$ was obtained from the fit Langevin equation described in Eq.~\ref{Eq1} (open symbols). The saturated moment per Fe follows a similar trend as $p_{eff}$ and $\theta_{CW}$ exhibiting a change in slope, at or around $x=0.44$. The $M(T)$ and $M(H)$ data measured at high magnetic fields data suggests that only samples with $x\leq0.3$ are FM ordered.

In Fig.~\ref{Figure3LowField}(a), $M(T)/H$ data measured at $H=20$~Oe are shown for all the  (Fe$_{1-x}$Ni$_x$)$_3$GeTe$_2$ samples. For $x\leq0.3$, the $M(T)$ curve shows a sharp increase in the magnetization at the temperatures corresponding to $T_C$ obtained from the zero-field resistance measurements [Fig.\ref{Figure3LowField}(b)].  The arrows indicate the peak in magnetization derivative d(M/H)/dT (not shown). In contrast, the $M(T)/H$ data for sample with $x \geq 0.37$  show a peak-like anomaly. The magnitude of the magnetization is significantly lower for samples with $x \geq 0.37$. The inset of Fig.~\ref{Figure3LowField}(a) shows a blow-up of the $M(T)/H$ data for the non-FM samples ($x \geq 0.37$) on a logarithmic scale. The peak-like anomaly at the magnetic transition (marked by arrows), $T_M$, persists for all samples up to $x=0.84$.
 
Figure~\ref{Figure3LowField}(b) shows the zero-field in-plane resistance, normalized by the resistance at $T=300~$K ($R_{300}$), for samples with $0\leq x\leq0.37$. All sample exhibit a low residual resistivity ratio, RRR $\approx  1$, which may be a result of the strong crystallographic site disorder exacerbated by Fe/Ni alloying and partial occupancy of the Fe/Ni2 site, which increases systematically with Ni substitution. Nevertheless, the signature of FM ordering, a kink in resistance associated with the loss of spin disorder scattering, is clearly evident in the resistance data for all samples with $x\leq 0.3$. Samples with $x>0.37$ become more insulating and the signature of the magnetic transition vanishes. The inset of Fig.~\ref{Figure3LowField}(b) shows the derivative of resistance (d\textit{R}/d\textit{T}), from which the FM transition temperature, $T_C$, is inferred. $T_C$ is marked by the arrows for each value of $x$, as the mid-point of the jump in d\textit{R}/d\textit{T}.

\subsection{Weak TF-$\mu$SR}

To gain microscopic insight into the evolution of magnetic order with Ni-substitution, muon spin relaxation ($\mu$SR) measurements (see Ref.~\onlinecite{Blundell_muSRreview} and Ref.~\onlinecite{Yaouanc_muSR} for a technical review) were performed on samples with different Ni compositions; two which have a FM ground state ($x =$ 0 and 0.3) and two beyond the apparent change in the nature of magnetic order  ($x=$0.56 and 0.76). The $\mu$SR technique relies on spin polarized, positive muons implanted in the sample. Once stopped inside the sample, their spin precesses around the local magnetic field \textbf{\textit{B}}, at the Larmor frequency, $\omega=\gamma_{\mu} B$,  where $\gamma_\mu$ is the gyromagnetic ratio of the muon. Muons decay with a lifetime $\tau =2.2$~$\mu$s, emitting a positron preferentially along the direction of the spin at the time of decay. Therefore, the measured asymmetry in positron counts at opposite sides of the sample, $A(t)$, is proportional to the muon spin polarization along this direction which reflects the local magnetic field distribution in the sample.  

\begin{figure}[t]
	\begin{center}
		\includegraphics[width=86mm]{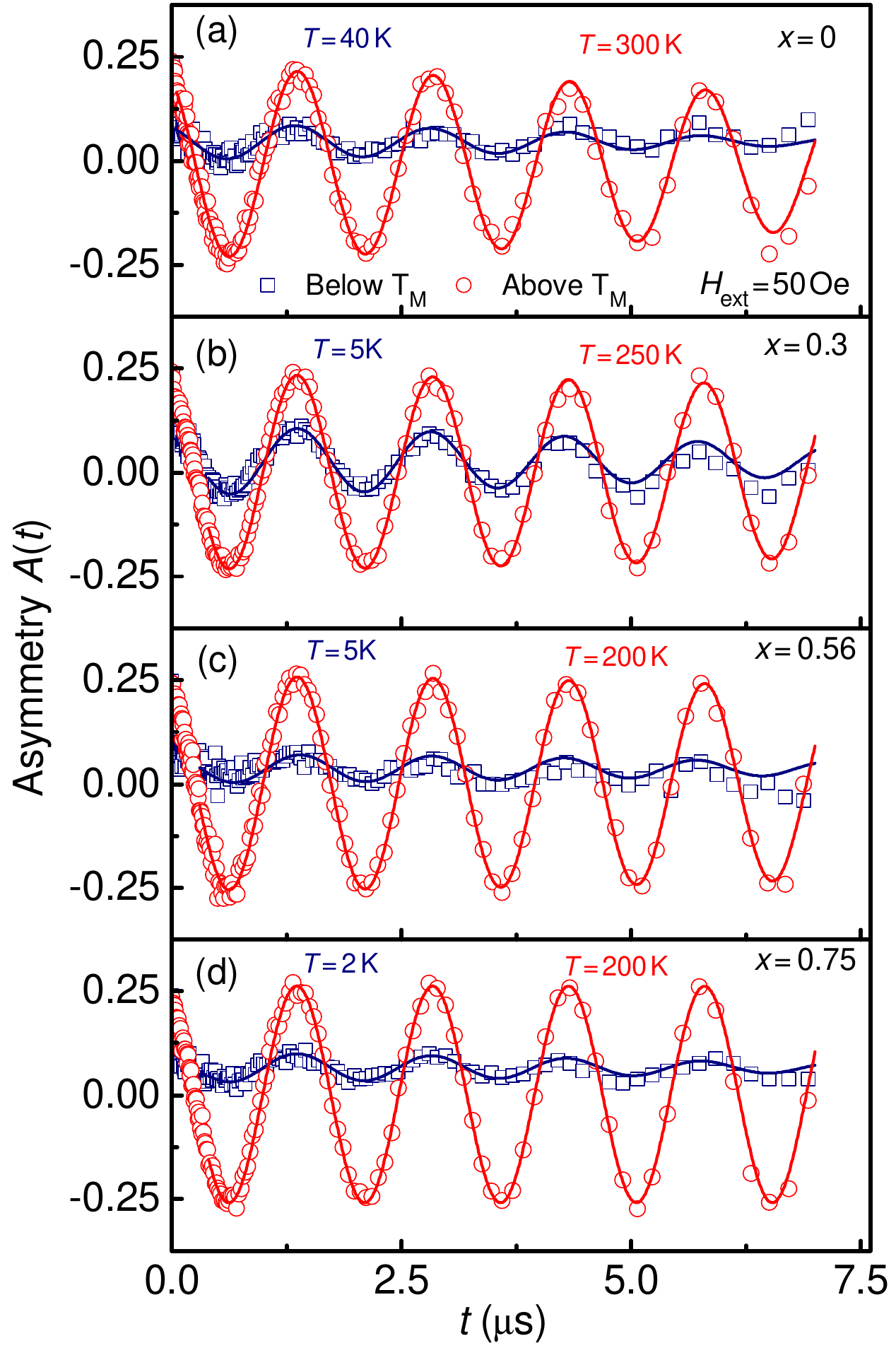}
	\end{center}
	\caption{TF-$\mu$SR asymmetry, $A\left( t \right)$, measured in a transverse eternal field  $H=50$~Oe, of selected (Fe$_{1-x}$Ni$_x$)$_3$GeTe$_2$ crystals with $x=0$ (a), 0.3 (b), 0.56 (c) and 0.75 (d) below (blue open squares) and above (red open circles) $T_M$. The solid lines are fit the model described in Eq.~\ref{eq2}. }
	\label{muSR1}

\end{figure}
\begin{figure}[t]
	\begin{center}
		\includegraphics[width=86mm]{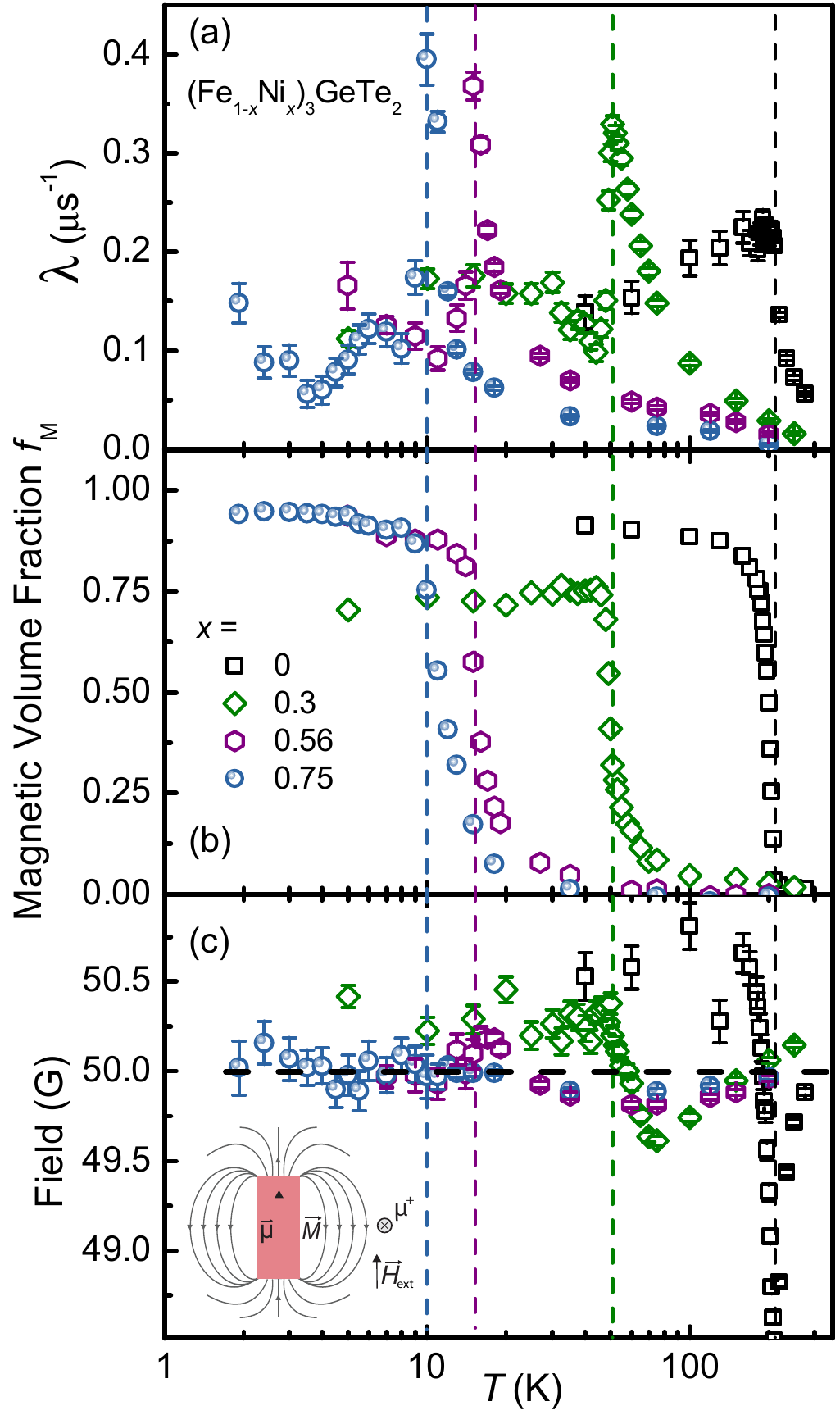}
	\end{center}
	\caption{(a) The muon relaxation, $\lambda$ vs. temperature, extracted from TF-$\mu$SR asymmetry fit to the model in Eq.~\ref{eq2}. (b) The magnetic volume fraction ${f_M}$ vs. temperature calculated from the ratio of the paramagnetic and magnetic amplitudes. (c) The local magnetic field experienced by muons stopped at the PM region of the sample. Note the negative field shift down from $H_\mathrm{ext} = 50$~Oe, which is apparent in all samples above the magnetic transition. The dashed lines indicate $T_M$~$\mu\rm{SR}$. Inset: Illustration of the fields sensed by the muons in proximity to a magnetic domain (pink) with a moment of \textbf{$\mu$}.}
	\label{muSR2}
\end{figure}

\begin{figure}[t]
	\begin{center}
		\includegraphics[width=86mm]{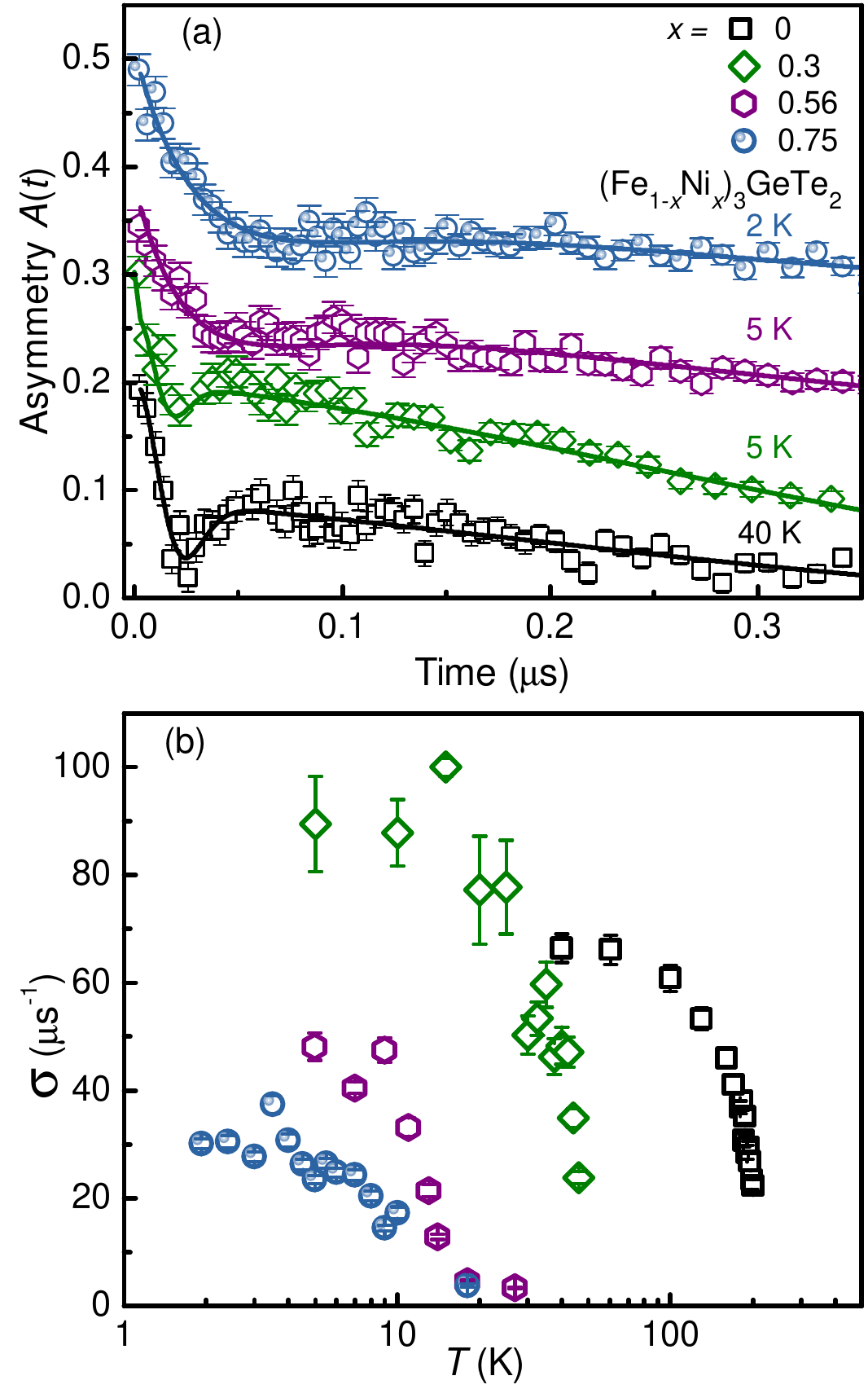}
	\end{center}
	\caption{(a) Early-time TF-$\mu$SR Asymmetry measured at the lowest \textit{T} for each sample, with a fit to the model in described in Eq~\ref{eq2}. The asymmetries were vertically stacked by 0.08 for clarity. (b) Static Gaussian field distribution width $\sigma$ at short times $\sigma$ obtained form the fit to Eq~\ref{eq1} dominated by the early-time asymmetry. }
	\label{muSR3}
\end{figure}

Typical TF-$\mu$SR spectra, measured in a weak transverse field of $H_\mathrm{ext}=50$~Oe, are presented in Fig.~\ref{muSR1}(a)-(d) for selected (Fe$_{1-x}$Ni$_x$)$_3$GeTe$_2$ crystals with $x=$ 0, 0.3, 0.56 and 0.75, below (blue open squares) and above (red open circles) their respective magnetic transition temperature $T_M$. Well above $T_M$, the oscillation in the asymmetry is weakly damped in all samples with a large amplitude ($\sim 0.24$), which represents the maximum amplitude measured in the GPS spectrometer.  This is clear evidence that at high temperature all samples are fully paramagnetic. In contrast, we observe a heavily damped and much smaller oscillating amplitude at low temperatures, which indicates that a large fraction of the samples is magnetic at these temperatures. In order to parametrize the behavior of the samples we fit the measured $A(t)$ in all samples, over the entire temperature range, to

\begin{equation}\label{eq1} 
A\left( t \right) = A_{PM}\left(T \right)P_{PM} + A_{M}\left(T \right)P_{M}
\label{eq2}
\end{equation}
where the sum of $A_{PM}$ (paramagnetic amplitude) and $A_{M}$ (magnetic amplitude) is determined by the experimental geometry and was fixed for each sample. The term $P_{PM}\left( t \right)  =\exp \left( { - \lambda t} \right)\cos \left( {{\gamma _\mu }B t + \phi } \right)$ describes the signal from muons stopping in paramagnetic regions of the sample and precess at the Larmor frequency in the local magnetic field, and $\lambda$ is the damping (relaxation) rate of the oscillating signal~\cite{Dalmas1997}. In the magnetically ordered regions, the local magnetic field is much larger than $H_{\mathrm{ext}}$. There, $A\left( t \right)$ can be described by a static Gaussian field distribution with a width $\sigma $. This leads to 
\begin{equation} 
P_M \left( {t} \right) = \frac{1}{3}+\frac{2}{3}(1-\sigma^2 t^2) \exp(-\sigma^2 t^2/2)
\end{equation}


i.e. a Gaussian Kubo-Toyabe depolarization function~\cite{Hayano1979} which is dominant at early muon decay times. The parameters extracted from best fits, for all samples, and as a function of temperature are summarized in Fig~\ref{muSR2}(a)-(c).

We start by discussing $\lambda$ as a function of temperature for the different samples shown in Figure~\ref{muSR2}(a). This parameter reflects the width of static field distribution present in the paramagnetic regions of the sample, $\Delta$, as well as the spin lattice relaxation rate, $1/T_1$, due to dynamic components in the local magnetic field experienced by the muons~\cite{Hayano1979}.  In all samples, $\lambda$ increases sharply as we approach $T_M$ from above, peaks at $T_M$ and then decreases and saturates below $T_M$. We define $T_M$~$\mu$SR as the peak temperature of $\lambda$.This is the typical behavior observed in systems undergoing a magnetic transition (for example, see Refs.~\onlinecite{Guidi2001, Krieger2017}). The transition temperature inferred form the TF-$\mu\rm{SR}$ ($T_M$~$\mu\rm{SR}$) measurements are indicated in Figure~\ref{muSR2} by the dashed vertical lines.

In Figure~\ref{muSR2}(b), the temperature dependence of the magnetic volume fraction is presented on a semi-logarithmic scale. This is calculated from the magnetic and paramagnetic amplitudes as ${f_M} = {{{A_M}} \mathord{\left/ {\vphantom {{{A_M}} {\left( {{A_M} + {A_{PM}}} \right)}}} \right.\kern-\nulldelimiterspace} {\left( {{A_M} + {A_{PM}}} \right)}}$ and represents the magnetic volume fraction i.e. regions where the implanted muons experience a broad distribution of local static fields resulting in a fast  depolarization due to incoherent precession. For the $x=0$ sample (black squares), ${f_M}$ sharply rises around $T=215$~K, which coincides with $T_{\mathrm{C}}$ determined from resistivity and magnetization measurements. The $x=0.3$ sample (green diamonds) shows a gradual rise of ${f_M}$, however it also sharply rises close to the FM ordering temperature. In contrast, the volume fraction of $x=0.56$ (purple hexagon) and $x=0.75$ (blue circles) sample shows a broader transition, however the sharp upturn concurs with the peak observed in bulk magnetization measurements [inset of Fig.~\ref{Figure3LowField}(a)]. 

For all samples, the magnetic volume fraction does not reach 100\%. This is partially a result of muons which stop in the sample holder and partially due to inclusion of remnant flux inter-grown between the (Fe$_{1-x}$Ni$_x$)$_3$GeTe$_2$ crystals/layers. The latter has been clearly observed in the $x=0.3$ set of crystals, post-measurement, which accounts for the $\sim25\%$ missing magnetic volume fraction. However, since the temperature dependence of the muon spin polarization comes predominantly from the magnetic regions in the sample, these inclusions do not affect the main conclusions drawn from these measurements.

We now turn to discussing the peculiar temperature dependence of the average local field, $B$, experienced by the muons shown in Fig~\ref{muSR2}(c). The field is extracted from the oscillating component of our measured signal ($P_{PM}$) and therefore reflects the size of the average field in the regions which have not ordered magnetically yet. In both $x=0$ and $x=0.3$ samples, a significant negative shift in $B$ is detected as the sample is cooled through $T_M$, followed by a sharp increase as \textit{T} is decreased further. The field saturates at lower temperatures at $B>H_{\rm{ext}}$. The temperature dependence in the $x=0.56$ and $x=0.75$ samples is dramatically different. In particular, the field shift above $T_M$ is smaller, but still negative and finite. Below $T_M$, the field saturates at $B = H_{\rm{ext}}$. A shift in the average local field from $H_{\rm{ext}}$ indicates a spontaneous magnetization in the sample, the difference between $x=0,0.3$ and $x=0.56,0.76$ samples, again suggests a different nature of magnetic ordering between low and high Ni content in this system.


Figure.~\ref{muSR3}(a) shows the early time behavior of the asymmetry (vertically shifted for clarity)  with fits to the model (solid lines) described in Eq~\ref{eq1}.  For $x=0$ (black squares), the spectra shows a strong dip which indicates consistent with developed FM order in the sample. The $x=0.3$ spectra  (green diamond), exhibits a shallow dip and then a slight recovery, as is typical when the muons experience a broader Gaussian distribution of local static fields. The $x=0.56$ and $0.76$ show only a quick decay of the asymmetry with a very broad and shallow dip which indicates smaller local static fields with an even broader distribution. For completeness, the values of $\sigma$, as a function of temperature for all samples are presented in Fig.~\ref{muSR3}(b). Above $T_M$, $\sigma=0$ since the full volume of the sample is paramagnetic, i.e. no static fields are sensed by the muons. Below $T_M$, $\sigma$ increases and saturates at low temperature when magnetic order is established in the sample. The saturation values of $\sigma$ for low doping are much higher than those measured in samples with higher Ni content, consistent with the smaller saturated moment and average moment per cluster or magnetic domain  observed in the magnetization data (Fig.~\ref{Figure2HighFied}).

\section{Discussion}
\label{Discussion}

\begin{figure}[t]
	\begin{center}
		\includegraphics[width=86mm]{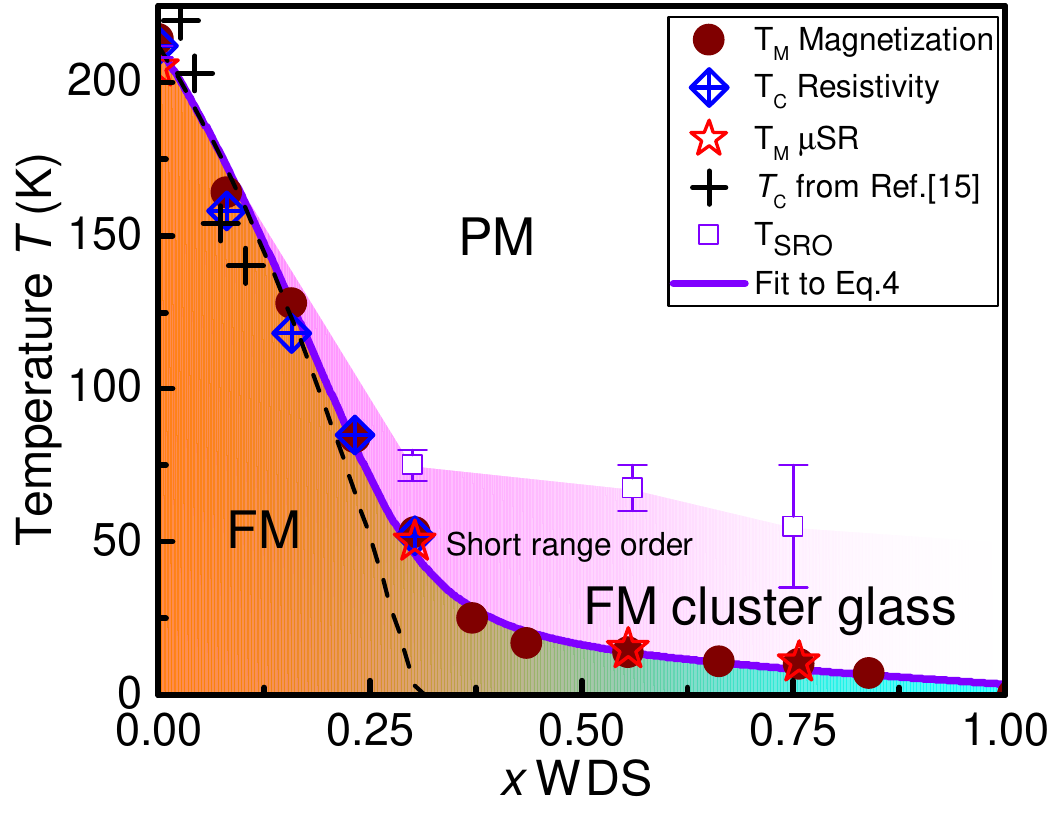}
	\end{center}
	\caption{The phase diagram of (Fe$_{1-x}$Ni$_x$)$_3$GeTe$_2$ determined from magnetization and TF-$\mu$SR ($T_M$), and resistivity measurements ($T_C$), showing a FM region up to $x=0.3$ which is smeared into a FM spin glass. The violet open squares represent $T_{\mathrm{SRO}}$ which was determined from the minumum in the Field at the muon site (Fig.~\ref{muSR2}(c)). The solid violet line denotes the fit of the magnetic ordering temperatures to a model described by Eq.~\ref{eq3}. The dashed line reflects only the classical dilution effect of disorder ($a=0$ in Eq.\ref{eq3}).}
	\label{PhaseDiagram}
\end{figure}

The magnetic properties of the (Fe$_{1-x}$Ni$_x$)$_3$GeTe$_2$ system vary significantly with Ni substitution. The parent compound Fe$_3$GeTe$_2$ clearly has a ferromagnetic ground state as been demonstrated in this, and previous studies~\cite{Deiseroth2006,Chen2013,May2016}. With the introduction of Ni to the system, $T_{\mathrm{M}}$, $p_{\mathrm{eff}}$, $\theta_{CW}$ and $M_{s}$ (Fig.~\ref{Figure2HighFied}) are suppressed, however bulk magnetic properties are consistent with long-range FM order which persist up to $x=0.3$. The inflection in slope of the inferred parameters [Fig.~\ref{Figure2HighFied}(c)-(e)], loss of the resistive anomaly (inset of Fig~\ref{Figure3LowField}(b)), change in the local field distribution as probed with weak TF-$\mu$SR (Fig.~\ref{muSR3}) and lack of a developed oscillation in the early-time $\mu$SR asymmetry, all indicate a dramatic change of the magnetic ground state in samples with $x>0.30$. In particular, the $M(H)$ data suggests a transition from FM order into a cluster glass state with a small moment per cluster and a nearly full magnetic volume fraction (Fig.\ref{muSR2}(b)). Note also the increase in the maximum value of $\lambda$ while $\sigma$ decreases with increasing Ni content. This indicates an enhanced dynamics in the local magnetic field near the magnetic transition accompanied by a decrease in the size of the local static fields.
  
The average local field [Fig.\ref{muSR2}(c)] experienced by muons stopped in the paramagnetic regions, can shed light on the nature of magnetic order in the two regimes. The large negative field shift observed for the $x=0$ and $0.3$ samples above $T_{\mathrm{C}}$  is indicative of the formation of FM regions, whose magnetic moment is aligned with the applied field. These regions produce a demagnetizing field which reduces the total magnetic field experienced by the muons stopping outside these regions (see illustration in the inset of Fig.~\ref{muSR2}(c)). As for samples with $x=0.56$ and $0.76$, the local field shift is significantly reduced compared to the FM samples, however, it remains finite and negative. This indicates that the magnetically ordered regions in these samples are either not aligned with the applied field, have a lower net magnetic moment or just smaller in size. Hence, they do not produce a large demagnetizing field outside the magnetic regions. 

Moreover, the minimum observed in the field shift in Fig.~\ref{muSR2}(c), occurs above $T_{\rm{M}}$ (marked by the dashed lines) for samples with $x\geq 0.3$. This suggest fluctuating clusters of short range magnetic order in the sample, which occur as a precursor to the long range order below $T_{\rm{M}}$,  or, for larger x, the formation of a ferromagnetic 
cluster glass.  We can define the temperature at which minimum in the field occurs as $T_{\rm{SRO}}$ (Short Range Order). In addition, one can rule out AFM interactions in the cluster glass phase, since AFM spin clusters would have a zero net moment which cannot produce demagnetizing fields observed as a negative field shift in Fig~\ref{muSR2}(c) in the $x=0.56$ and $0.76$ data. Therefore, all these observations support a cluster glass state with FM interactions for Ni concentrations above $x=0.3$. It is worth noting that any other short range correlations will not produce a negative field shift. For example AFM or a random spin glass will produce clusters with zero net magnetic moment and therefore not shift in the precession frequency.

Another aspect to consider is the role of the structural difference between the end members, Fe$_3$GeTe$_2$ and Ni$_3$GeTe$_2$, on the magnetic ground state. One might speculate that a structural transition between the two structure types can drives the observed change of magnetic order. However, the \textit{c}-lattice parameter (Fig.~\ref{Figure1Stucture}(d)) and site occupancies (Fig.~\ref{Figure1Stucture}(e)) continuously change across the Ni-composition range. There is no symmetry change, only continuous changes in site occupancy. It is therefore unlikely that a structural transition occurs when continuously going from the Fe$_3$GeTe$_2$ to the Ni$_3$GeTe$_2$ prototype. 

Figure~\ref{PhaseDiagram} summarizes the ordering temperatures of the (Fe$_{1-x}$Ni$_x$)$_3$GeTe$_2$ system inferred from  zero field resistivity ($T_{\mathrm{C}}$) (blue diamond), low field magnetization data ($T_{\mathrm{M}}$) (dark red circles) and TF-$\mu$SR measurements (red stars).  As Fe is substituted for Ni in this system, long range FM order is suppressed down from $T_C=212$~K down to $T_C=52$~K for $x=0.3$. Above $x=0.37$, long range FM order is continuously smeared into a glassy magnetic phase, below $T_{\mathrm{M}}$, which persists up to $x=0.84$ (and possibly higher Ni concentrations). Short range magnetic order persists in the temperature range between $T_{\mathrm{M}}$ and $T_{\mathrm{SRO}}$ (violet open squares) for $x\geq 0.3$. Moreover, Ni substitution suppresses FM order in Fe$_3$GeTe$_2$ equivalently to Fe vacancies (black crosses)~\cite{May2016}.

The (Fe$_{1-x}$Ni$_x$)$_3$GeTe$_2$ system exhibits a typical behavior of strongly disordered ferromagnets where long range order is smeared into a glassy phase~\cite{Brando2016}. In these  systems, the shape of the phase diagram can be qualitatively described by two competing effects. One is a classical dilution effect that suppresses $T_{\mathrm{C}}$ to zero at sufficiently large values of \textit{x}, where  $x \propto 1/\tau$ is a dimensionless measure of the disorder and $\tau$ is the elastic mean free path~\cite{Cardy1996}. This can be generally expressed as $\frac{{T_C}\left( x \right)}{{T_C}\left( 0 \right)} =  1 - x - x^2$. However, at sufficiently low temperature the diffusive motion of the electrons  increases in the effective exchange interaction, which can enhanced $T_{\mathrm{C}}$~\cite{Altshuler1983}. This effect is linear for small disorder at $T = 0$, and is strongest for small values of $T_{\mathrm{C}}$. Assuming that the disorder is proportional to the Ni content, one can substitute $x \rightarrow s x$, where $s$ is the scaling factor between  them. A simple schematic way to represent both effects is 

\begin{equation} 
\frac{{T_C}\left( x \right)}{{T_\mathrm{C}}\left( 0 \right)} =  1 - sx - {s^2x^2} + \frac{{asx}}{{1 + b{T_c}\left( x \right)}/sx}.~\cite{Brando2016}
\label{eq3}
\end{equation}
Here \textit{a} and \textit{b} signify the strength and cutoff of the effect.  The solid violet line shown Fig.~\ref{PhaseDiagram} denotes the fit of the magnetic ordering temperatures to Eq.\ref{eq3} where $a$ was fixed to 1 (if allowed vary freely, $a=0.8 \pm 0.4$ with a negligible difference on the other fit parameters), $b = 23\pm 4$, $s=2.02\pm0.05$. The dashed line in Fig.~\ref{PhaseDiagram} reflects the classical dilution effect of disorder ($a=0$ in Eq.\ref{eq3}) . The critical Ni concentration can be inferred from the fit value of \textit{s}, which yields $x_c = 0.31 \pm 0.01$, which is consistent with the observation of long range FM order vanishing above the inflection point around $x=0.3$. 
 
\section{Conclusion}
\label{Conclusion}

In summary, we studied the effect of Ni substitution on the structural properties and the FM ground state of single crystalline samples of (Fe$_{1-x}$Ni$_x$)$_3$GeTe$_2$ with $x = 0-0.84$. Single crystal X-ray diffraction and refinement have shown that Fe can be continuously substituted with Ni without significant structural variations. Magnetization and resistivity measurements have shown that Ni suppresses FM order from $T_\mathrm{C}$=212~K for $x=0$ down to $T_\mathrm{C}$=50~K for $x=0.3$, as well as a strong suppression of M$_s$, $p_{\mathrm{eff}}$, and $\theta_{\mathrm{CW}}$. We also find that Ni suppresses FM order in a similar fashion to Fe deficiencies in Fe$_{3-x}$GeTe$_2$. TF-$\mu$SR measurements have revealed that for $x>0.3$ FM order is continuously smeared into a FM cluster-glass phase, with a nearly full magnetic volume fraction.  

\section*{Acknowledgment}

Work done at Ames Laboratory was supported by US Department of Energy, Basic Energy Sciences, Division of Materials Sciences and Engineering under Contract NO. DE-AC02-07CH111358.  G.D. was funded by the Gordon and Betty Moore Foundation’s EPiQS Initiative through Grant GBMF4411. Part of this work is based on experiments performed at the Swiss Muon Source, S$\mu$S, Paul Scherrer Institute, Villigen, Switzerland.

\bibliography{Bibliography}

\end{document}